\documentclass[aps,reprint,amsmath,amssymb,showkeys,twocolumn]{revtex4-2}
\usepackage{graphicx}
\usepackage{amscd}
\usepackage{dcolumn}
\usepackage{bm}
\begin{document}

\title{Note on cosmographic approach to determining parameters of Barrow entropic dark energy model}

\author{Yu.L. Bolotin}
\affiliation{National Science Center Kharkov Institute of Physics and Technology, 1 Akademicheskaya str.,  Kharkov, 61108, Ukraine}
\author{V.V.Yanovsky}
\email{yanovsky@isc.kharkov.ua}
\affiliation{V. N. Karazin Kharkiv National University, 4 Svobody Sq., Kharkiv, 61022, Ukraine}
\affiliation{Institute for Single Crystals, NAS Ukraine, 60 Nauky Ave., Kharkov, 31001, Ukraine}

\date{\today}

\begin{abstract}
The cosmographic approach is used to determine the parameters of the Barrow entropic dark energy model. The model parameters are expressed through the current kinematic characteristics of Universe expansion.
\end{abstract}


\maketitle

1.The viability (efficiency) of any cosmological model is determined by  its ability to reproduce the results of observations. A mandatory preliminary stage for testing efficiency is determining the model parameters.

There are two approaches to determining the parameters of cosmological models.  The first, extremely time-consuming approach consists of -- sequentially exploring of the parametric space in order to find the optimal set of parameters. The lack of unambiguous criteria for the concept of ``optimal set'' (especially in the case of multipara meter models) and the need to use significant computing resources reduce the effectiveness of this approach. Nevertheless, ``blind'' search of parameters remains the dominant method for finding parameters of cosmological models.

Let us now formulate an alternative approach for finding modelparameters \cite{bolotin2018applied}. Instead of searching through parameters in order to find the optimal set, let's build a procedure that will allow us to relate the observed characteristics of the evolution of the Universe with the parameters of the cosmological model used to describe this evolution. That is, we change the direction of movement: not from parameters to observations, but from observations to parameters. The advantages of this approach will be discussed in detail later.

The first step towards this goal is to select observables that will be used to find the parameters of cosmological models. The main requirement for this set is the maximum possible modellessness.

The required modellessness can be achieved if the kinematic characteristics of the expansion of the Universe or, in other words, cosmographic parameters are chosen as the initial set of observables \cite{Visser_2005,Visser_2004}. Cosmography represents the kinematics of cosmological expansion \cite{Weinberg:1972kfs}, and cosmographic parameters are the coefficients of the Taylor series expansion of the scale factor $a(t)$.

In the early 70s of the twentieth century, Alan Sandage \cite{1970PhT_1970} defined as the main goal of cosmology  the determination of two parameters: the Hubble parameter and the deceleration  parameter. Everything seemed simple and clear: the Hubble parameter determines the expansion rate of the Universe, and the deceleration parameter takes into account small corrections due to the decrease in the expansion rate due to gravity. However, the situation turned out to be much more complicated than expected.

For a more complete description of the kinematics of cosmological expansion, it is useful to consider higher order derivatives of the scale factor \cite{bolotin2018applied,Bolotin_2012,Dunsby_2016}
\begin{equation} \label{eq1}
\begin{array}{l} {H(t)\equiv \frac{1}{a} \frac{da}{dt} ;} \\ {q(t)\equiv -\frac{1}{a} \frac{d^{2} a}{dt^{2} } \left[\frac{1}{a} \frac{da}{dt} \right]^{-2} ;} \\ {j(t)\equiv \frac{1}{a} \frac{d^{3} a}{dt^{3} } \left[\frac{1}{a} \frac{da}{dt} \right]^{-3} ;} \\ {s(t)\equiv \frac{1}{a} \frac{d^{4} a}{dt^{4} } \left[\frac{1}{a} \frac{da}{dt} \right]^{-4} ;} \\ {l(t)\equiv \frac{1}{a} \frac{d^{5} a}{dt^{5} } \left[\frac{1}{a} \frac{da}{dt} \right]^{-5} } \end{array}
\end{equation}
The parameters of any model that satisfies the cosmological principle can be expressed through a set of cosmographic parameters \eqref{eq1}.

2. The goal of this work is to determine the parameters of holographic dark energy based on the Barrow entropy \cite{Barrow_2020}.
\begin{equation} \label{eq2}
S_{B} =\left(\frac{A}{A_{0} } \right)^{1+\frac{\Delta }{2} }
\end{equation}
where $A$ is the area of the horizon, $A_{0}$ is the Planck area, and $\Delta $ is a free parameter of the model, $0\le \Delta \le 1$. The choice of entropy is dictated by the desire to take into account quantum deformations of the horizon surface. The measure of this deformation leading to the fractal structure of the horizon is the new parameter $\Delta$, which takes the value $\Delta =0$ in the  undeformed case of Bekenstein entropy, and the value $\Delta =1$ corresponds to the maximum deformation leading to an increase in the fractal dimension of the horizon surface by one. Barrow entropy is a fractal generalization of Bekenstein entropy \cite{Bekenstein:1972tm,PhysRevD.7.2333}.

Density of entropic dark energy generated by entropy \eqref{eq2} \cite{Wang_2017,LI20041,Saridakis_2008,Saridakis_2020,anagnostopoulos_2020,Srivastava_2020,Saridakis_2020b}
\begin{equation} \label{eq3}
\rho _{e} =CL^{\Delta -2}
\end{equation}
where $C$ is the dimensional parameter $[C]=L^{-\Delta -2}$ and $L$ is the infrared cutoff scale. At $\Delta =0$, when the Barrow entropy reduces to the Bekenstein entropy, the Barrow entropicy dark energy (BEDE) density reduces to the standard holographic energy density $\rho _{e} {\rm \sim }L^{-2}$. Choosing the Hubble radius $H^{-1} $ as the infrared macro scale, we obtain
\begin{equation} \label{eq4}
\rho _{e} =CH^{2-\Delta }
\end{equation}
The evolution of a flat FLRW Universe filled with nonrelativistic matter $\left(\rho _{m} ,p_{m} =0\right)$ and entropic dark energy $\left(\rho _{e} ,p_{e} \right)$ is described by a system of equations
\begin{equation} \label{eq5}
H^{2} =\frac{8\pi G}{3} (\rho _{m} +\rho _{e} )
\end{equation}
\begin{equation} \label{eq6}
\dot{\rho }_{m} +3H\rho _{m} =0
\end{equation}
\begin{equation} \label{eq7}
\dot{\rho }_{e} +3H(\rho _{e} +p_{e} )=0
\end{equation}

3. In 2008 Dunaisky and Gibbons \cite{Dunajski_2008} proposed an original approach to determining the parameters of models that satisfy the cosmological principle. Let's briefly describe the essence of the proposed approach.  Consider the $n$-parametric cosmological model. Assuming that the dynamical variables are multiply differentiable functions of time, let us differentiate the initial evolution equation $n$ times. The obtained system will be used to express the free parameters in terms of time derivatives of the dynamical variables. For cosmological models based on the FLRW metric, model parameters can be expressed in terms of time derivatives of the Hubble parameter. These derivatives are directly related to cosmographic parameters
\begin{equation} \label{eq8}
\begin{array}{l} {\dot{H}=-H^{2} (1+q);} \\ {\ddot{H}=H^{3} \left(j+3q+2\right);} \\ {\dddot{H}=H^{4} \left[s-4j-3q(q+4)-6\right];} \\ {\stackrel{....}{H}=H^{5} \left[l-5s+10\left(q+2\right)j+30(q+2)q+24\right]} \end{array}
\end{equation}
Note that the Friedman equation for a specific model can  be presented in terms of cosmographic parameters. For example, in SCM the Friedman equation takes the form
\begin{equation} \label{eq9}
s+2(q+j)+qj=0
\end{equation}
By repeatedly differentiating of this, type of equation with respect to time and using
\begin{equation} \label{eq10}
\begin{array}{l} {\frac{dq}{dt} =-H\left(j-2q^{2} -q\right),} \\ {\frac{dj}{dt} =H\left[s+j\left(2+3q\right)\right],} \\ {\frac{ds}{dt} =H\left[l+s\left(3+4q\right)\right],} \\ {\frac{dl}{dt} =H\left[m+l\left(4+5q\right)\right]} \end{array}
\end{equation}
it is possible to express higher cosmographic parameters through a fixed set of lower parameters, known with sufficient accuracy from observations.

The cosmographic approach to determining model parameters has a number of advantages. To find model parameters, it is necessary to solve a system of algebraic rather than differential equations. The obtained relations between model and cosmographic parameters are exact. The approach under consideration can be effectively used for cosmological models with interaction in the dark sector \cite{Bolotin_2015}.

The proposed approach may be especially useful in entropy cosmology. The point is that we should first find out which classes of entropy forces (or models with fixed entropy forces) are suitable for describing cosmological evolution. The decisive role in answering this question is played by the existing (or future) results of cosmological observations that determine the parameters of the models. Expressing model parameters through cosmographic parameters automatically solves the problem of matching model parameters and observations that determine the model parameters.

4. Let us apply the procedure described above to the two-parametric BEDE model
\begin{equation} \label{eq11}
\begin{array}{l} {\rho _{e} =CL^{\Delta -2} =CH^{2-\Delta } =CH^{\gamma } ,} \\ {\gamma \equiv 2-\Delta } \end{array}
\end{equation}
The Friedman equation \eqref{eq7} and conservation equations for each component \eqref{eq8}, \eqref{eq9} allow us to obtain a system of equations for determining model parameters
\begin{equation} \label{eq12}
\begin{array}{l} {\dot{H}+\frac{3}{2} H^{2} =\frac{3}{2} f,} \\ {\ddot{H}+3H\dot{H}=\frac{3}{2} \gamma \frac{\dot{H}}{H} f,} \\ {f\equiv \frac{1}{3} \rho _{e} } \end{array}
\end{equation}

Solutions of this system
\begin{equation} \label{eq13}
\begin{array}{l} {\gamma =\frac{\left(H\ddot{H}+3H^{2} \dot{H}\right)}{\dot{H}\left(\dot{H}+\frac{3}{2} H^{2} \right)} =\frac{\left(\frac{\ddot{H}}{\dot{H}H} +3\right)}{\frac{\dot{H}}{H^{2} } +\frac{3}{2} } } \\ {C=H^{-\gamma } \left(2\dot{H}+3H^{2} \right)} \end{array}
\end{equation}
allow, using time derivatives of the Hubble parameter \eqref{eq8}, to express model parameters in terms of cosmographic parameters
\begin{equation} \label{eq14}
\begin{array}{l} {\Delta =2\left[\frac{\left(1-j\right)}{\left(1+q\right)\left(1-2q\right)} +1\right],} \\ {C=\frac{1-2q}{H^{\gamma -2} } } \end{array}
\end{equation}
Model parameters $\Delta $ and $C$ are constants, and cosmographic parameters, generally speaking, are functions of time. However, it can be shown \cite{bolotin2018applied} that $\frac{d\Delta }{dt} =0$ and$\frac{dC}{dt} =0$. Therefore, the right-hand sides in \eqref{eq14} can be calculated at any moment in time. Using cosmografphc parameters $H_{0} ,q_{0} ,j_{0} $ at the current moment in time, we present \eqref{eq14} in the form
\begin{equation} \label{eq15}
\begin{array}{l} {\Delta =2\left[\frac{\left(1-j_{0} \right)}{\left(1+q_{0} \right)\left(1-2q_{0} \right)} +1\right],} \\ {C=\frac{1-2q_{0} }{H_{0} ^{\gamma -2} } } \end{array}
\end{equation}
Relations \eqref{eq15} solve the problem of finding the parameters of the BEDE model for a fixed set of cosmographic parameters. We emphasize that relations \eqref{eq15} are exact. Therefore, the uncertainty of model parameters is associated only with the current inaccuracy of the values of cosmographic parameters. At present, the current value of the deceleration parameter $q_{0} $is known quite well \cite{Mukherjee_2022}. Depending on the form of parameterization $q(z)$, the current value of the deceleration parameter $q_{0}$ takes on the values

\[ q_0 \, -0.573^{+0.041}_{-0.042}  \, -0.580^{+0.055}_{-0.063}  \, -0.533_{-0.038}^{+0.038}  \, -0.574_{-0.045}^{+0.044} \]

This result is in good agreement with the current value of the deceleration parameter in the SCM
\begin{equation} \label{eq16}
q_{0} =\frac{1-3\Omega _{\Lambda 0} }{2} \simeq 0.6
\end{equation}
where $\Omega _{\Lambda 0} $ is current relative density of dark energy in the form of a cosmological constant.  The situation with determining the current value of the jerk parameter $j_{0}$ is much worse \cite{Mamon_2018b,Zhai_2013}. Uncertainties reach about 100\%. However, increasing the accuracy of determining the parameter is only a matter of time. In particular, cosmography with next-generation gravitational wave detectors \cite{chen2024cosmography} will significantly improve the accuracy of determining cosmographic parameters. Let us recall that the first measurements of the Hubble parameter  differed by an order of magnitude from the present ones. Achieving the required accuracy will solve the problem of finding the parameters of two-parametric cosmological models. For models with a large number of parameters, higher cosmographic parameters will be required.

Since the current value of the deceleration parameter is well defined, we can, for a fixed value $q_{0} \left(q_{0} \simeq 0.6\right)$, determine the range of parameter values $j_{0}$ realized in the BEDE model at $0\le \Delta \le 1$ . From \eqref{eq15} we find
\begin{equation} \label{eq17}
j_{0} =-0.44\Delta +1.88
\end{equation}
For the limiting values of the model parameter $\Delta =0,1,2$ (SCM), we obtain
\begin{equation} \label{eq18}
\begin{array}{l} {\Delta =0\to j_{0} =1.88,} \\ {\Delta =1\to j_{0} =1.44} \\ {\Delta =2\to j_{0} =1} \end{array}
\end{equation}
The range of values $1.44\le j_{0} \le 1.88$ is not excluded by current observations $j_{0} =2,84_{-1.23}^{+1.00} $ \cite{capozziello2023}.

In SCM $j=1$, therefore $\Delta =2$, that goes beyond the permissible values $\Delta$. In SCM, dark energy is realized in the form of a cosmological constant, and is naturally $\rho _{e} =CH^{2-\Delta }$ transformed into a constant. Note that for $\Delta =2$ entropy (it can no longer be called Barrow entropy, for which $0\le \Delta \le 1$) $S_{t4} =\left(\frac{A}{A_{0} } \right)^{1+\frac{\Delta }{2} } \propto r_{H}^{4}$. Thus, values $\Delta =0,1,2$ generate entropies proportional to $r_{H}^{2}$ (Bekenstein entropy), $r_{H}^{3}$ and $r_{H}^{4}$ respectively. The corresponding entropic forces lead to  appearance of an additional driving term $H^{2}, H$ and constant in the Friedman equation. These terms can be interpreted as the entropic dark energy density \cite{Komatsu_2016,PhysRevD123516}.

Recently, Basilakos et al \cite{basilakos2023barrow} considered BEDE  model with time-dependent parameters. This made it possible to vary the role of quantum effects at different stages evolution of the Universe. A first investigation of holographic dark energy models with  varying exponent was performed in \cite{Nojiri_2019,Nojiri_2022} and it was shown that the running behavior can lead to interesting physical results. In this work, we restrict ourselves to the BEDE model with time-independent parameters.
\begin{figure}
	\includegraphics[width=5 cm]{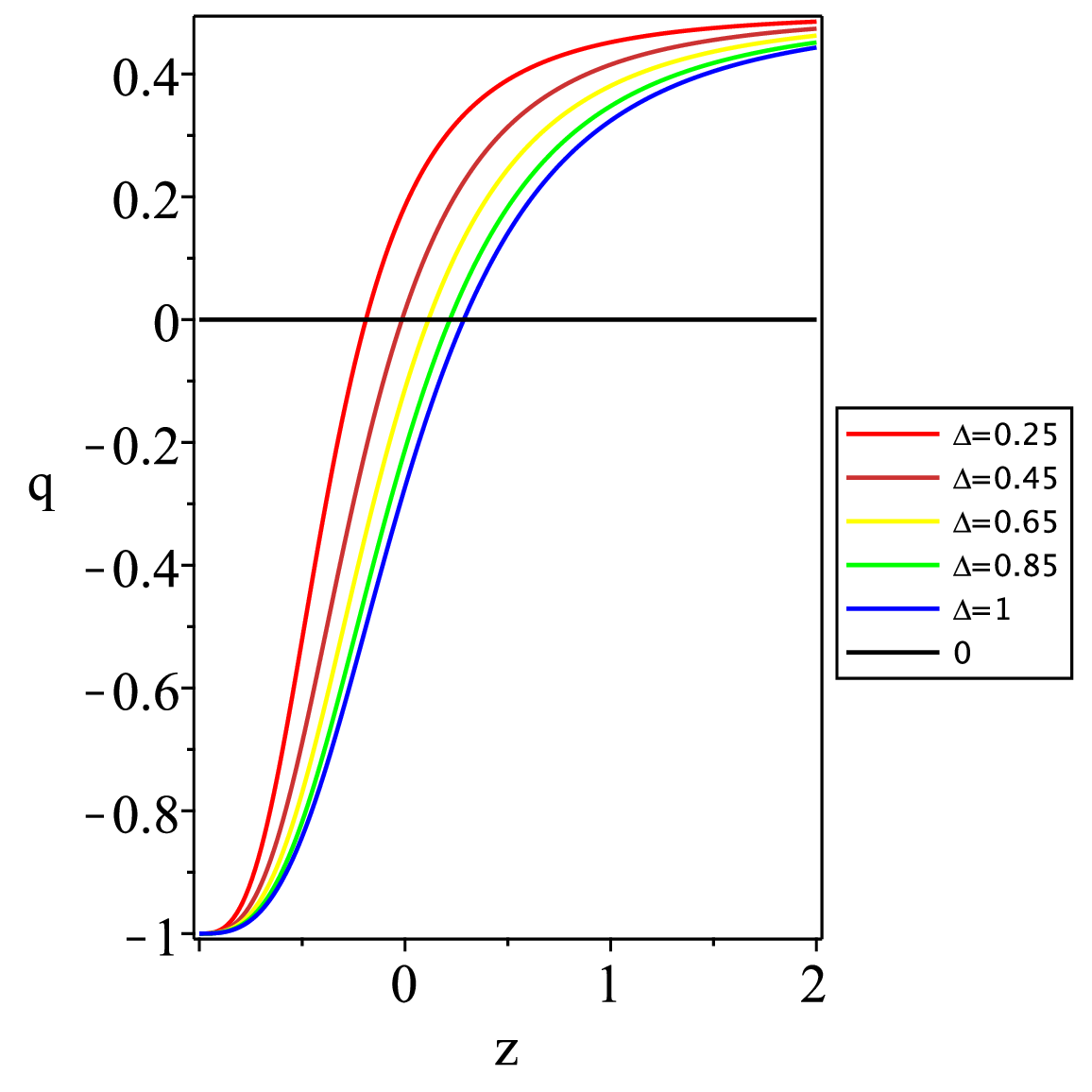}
	\caption{Evolution of the deceleration parameter as a function of redshift}
	\label{fg1}
\end{figure}

5. Let us now move on to the analysis of the relations connecting the deceleration parameter $q$and the model parameter $\Delta$. For a flat FLRW Universe filled with nonrelativistic matter and BEDE with relative densities $\Omega _{m}$ and $\Omega _{e}$ the deceleration parameter is equal to \cite{Pradhan_2021,dixit2021barrow}
\begin{equation} \label{eq19}
q=-\frac{\dot{H}}{H^{2} } -1=\frac{\left(\Delta +1\right)\Omega _{m} -\Delta }{\Delta -\left(\Delta -2\right)\Omega _{m} }
\end{equation}
where
\begin{equation} \label{eq20}
\begin{array}{l} {\Omega _{m} =\frac{\Omega _{m0} \left(1+z\right)^{3} }{\frac{\Omega _{m0} \left(1+z\right)^{3} +\Omega _{e0} }{} } ,} \\ {\Omega _{e} =\frac{\Omega _{e0} }{\frac{\Omega _{m0} \left(1+z\right)^{3} +\Omega _{e0} }{} } ,} \\ {\Omega _{m} +\Omega _{e} =1} \end{array}
\end{equation}
For a Universe of arbitrary curvature \cite{dixit2021barrow}
\begin{equation} \label{eq21}
q=\frac{\left(\Delta +1\right)\Omega _{m} -\Delta \left(\Omega _{k} +1\right)}{\Delta +\left(\Delta -2\right)\Omega _{k} -\left(\Delta -2\right)\Omega _{m} } ,
\end{equation}
where $\Omega _{k} =\frac{k}{a^{2} H^{2} } $.

Expression \eqref{eq19} for the deceleration parameter has the correct asymptotics, independent of $\Delta $. In the early Universe $\Omega _{m} =1(z\to \infty )\; q=\frac{1}{2} $, and in the distant future $\Omega _{m} =0\left(z=-1\right)\quad q=-1$.

The evolution of the deceleration parameter $q$ as a function of redshift$z$, described by expression \eqref{eq19} is shown in Fig.~\ref{fg1}

Fig.\ref{fg1} shows that the BEDE model can well explain the history of the Universe, including the era of matter dominance.

As we noted above, this approach represents a ``blind'' search of parameters in order to find the optimal value of the parameter$\Delta $. The optimality criterion can, for example, be the prediction of a fairly well-known value of redshift$\left(z_{q=0} \right)$, at which the transition from deceleration to accelerated expansion occurs, i.e. the deceleration parameter changes sign \cite{dixit2021barrow,Farooq_2013,Farooq_2017,Mamon_2018c,Mamon-2018d,Mamon_2016b,Maga_a_2014}. From \eqref{eq19} we find
\begin{equation} \label{eq22}
q=0\to \Delta =\frac{\Omega _{m} \left(z_{q=0} \right)}{1-\Omega _{m} \left(z_{q=0} \right)}
\end{equation}
From \eqref{eq22} follows that
\begin{equation} \label{eq23}
z_{q=0} =\left(\frac{\Delta \Omega _{e0} }{\Omega _{m0} } \right)^{1/3} -1
\end{equation}
For the value of the parameter $\Delta =2$ (in this case, the density BEDE is independent of time), expression \eqref{eq22} reproduces the transition acceleration in the SCM

The maximum transition redshift in the BEDE model is achieved at a degree of fractality $\Delta =1$ and equal to$z_{q=0}^{\max } \simeq {\rm 0.286}$. Coordination of the values of transition acceleration in the BEDE model with the corresponding value in the SCM requires going beyond the permissible values of the fractality parameter$0\le \Delta \le 1$.

6.Thus, we see that although the BEDE model generally correctly describes the evolution of the FLRW   Universe, it leads to a too late transition from  deceleration to accelerated expansion $\left(z_{q=0}^{BEDE} <z_{q=0}^{SCM} \right)$. One possibility to resolve this contradiction is to include interaction between dark matter and BEDE  \cite{Mamon_2021,LUCIANO2023101256}. The cosmographic approach to determining the parameters of models involving such interactions will be the subject of further research.

\bibliography{bibli2}

\end{document}